\font\twelverm=cmr10  scaled 1200   \font\twelvei=cmmi10  scaled 1200
\font\twelvesy=cmsy10 scaled 1200   \font\twelveex=cmex10 scaled 1200
\font\twelvebf=cmbx10 scaled 1200   \font\twelvesl=cmsl10 scaled 1200
\font\twelvett=cmtt10 scaled 1200   \font\twelveit=cmti10 scaled 1200
\font\twelvesc=cmcsc10 scaled 1200  \font\twelvesf=cmss10 scaled 1200
\skewchar\twelvei='177   \skewchar\twelvesy='60


\def\twelvepoint{\normalbaselineskip=12.4pt plus 0.1pt minus 0.1pt
  \abovedisplayskip 12.4pt plus 3pt minus 9pt
  \belowdisplayskip 12.4pt plus 3pt minus 9pt
  \abovedisplayshortskip 0pt plus 3pt
  \belowdisplayshortskip 7.2pt plus 3pt minus 4pt
  \smallskipamount=3.6pt plus1.2pt minus1.2pt
  \medskipamount=7.2pt plus2.4pt minus2.4pt
  \bigskipamount=14.4pt plus4.8pt minus4.8pt
  \def\rm{\fam0\twelverm}          \def\it{\fam\itfam\twelveit}%
  \def\sl{\fam\slfam\twelvesl}     \def\bf{\fam\bffam\twelvebf}%
  \def\mit{\fam 1}                 \def\cal{\fam 2}%
  \def\sc{\twelvesc}               \def\tt{\twelvett}
  \def\sf{\twelvesf}
  \textfont0=\twelverm   \scriptfont0=\tenrm   \scriptscriptfont0=\sevenrm
  \textfont1=\twelvei    \scriptfont1=\teni    \scriptscriptfont1=\seveni
  \textfont2=\twelvesy   \scriptfont2=\tensy   \scriptscriptfont2=\sevensy
  \textfont3=\twelveex   \scriptfont3=\twelveex  \scriptscriptfont3=\twelveex
  \textfont\itfam=\twelveit
  \textfont\slfam=\twelvesl
  \textfont\bffam=\twelvebf \scriptfont\bffam=\tenbf
  \scriptscriptfont\bffam=\sevenbf
  \normalbaselines\rm}


\def\singlespace{\baselineskip=\normalbaselineskip}

\def\doublespace{\baselineskip=\normalbaselineskip \multiply\baselineskip by 2}

\newcount\firstpageno
\firstpageno=2
\footline={\ifnum\pageno<\firstpageno{\hfil}\else{\hfil\twelverm\folio\hfil}\fi}
\def\toppageno{\global\footline={\hfil}\global\headline
  ={\ifnum\pageno<\firstpageno{\hfil}\else{\hfil\twelverm\folio\hfil}\fi}}
\let\rawfootnote=\footnote              
\def\footnote#1#2{{\rm\singlespace\parindent=0pt\parskip=0pt
  \rawfootnote{#1}{#2\hfill\vrule height 0pt depth 6pt width 0pt}}}


\hsize=6.5truein
\hoffset=0truein
\vsize=9.2truein
\voffset=-.2truein
\parskip=\medskipamount
\def\\{\cr}
\twelvepoint            
\doublespace            
\overfullrule=0pt       

\def\^#1{^{{}^{#1}}}
\def\_#1{_{{}_{#1}}}
\def\ssigma{\mathrel{\hbox{$\sigma$\kern-.6em\hbox{$/$}}}}
\def\sg#1{\mathrel{\hbox{$#1$\kern-.5em\hbox{$/$}}}}
\def\spartial{\mathrel{\hbox{$\partial$\kern-.5em\hbox{$/$}}}}

\def\A{\mathrel{\hbox{A\kern-.55em\hbox{$/$}}}}

\def\b'#1{\bar #1^{\,\prime}}
\def\approaches#1{\raise.6ex\hbox{${\mathop{\relbar\joinrel\relbar\joinrel
\rightarrow}\limits_{#1}}$}}

\input psfig

\singlespace
\centerline{\bf One loop vacuum polarization without infinities}
\centerline{A. A. Broyles}
\centerline{\it Department of Physics, Institute for Fundamental
Theory}
\centerline{\it University of Florida, Gainesville, FL 32611-8440}
\vskip 1cm
\noindent
A technique for avoiding infinite integrals in the calculation of
the one-loop diagram contribution to the vacuum polarization
component of an atomic energy level is presented. This makes
renormalization unnecessary. Infinite integrals do not occur
because, as it is shown, no delta functions are required for the
Green's functions. Thus there are none to overlap. This procedure
is shown to produce the same formula as the one obtained by
dimensional renormalization.

\vskip 1cm
\noindent {\bf I. INTRODUCTION}
\vskip .4cm
Divergent integrals appear in the calculations of the Lamb shift in
a number of papers and text books. They are removed by one of the
processes called ``renormalization''. The present author once asked
Richard Feynman if these infinite integrals represent a defect in
the fundamental equations of QED or do they originate in the method
used to solve the equations. He paused for a moment and then
replied, ``I don't know''. The same question to Paul Dirac produced
the answer, ``The basic equations are at fault.'' We shall see
evidence here, however, that there may be no defect in the
fundamental equations, but that the infinities may arise from the
use of faulty Green's functions as they do in many one-loop
calculations.

Actually, this is a solution to a very old problem that held up the
development of quantum electrodynamics (QED) in the years preceding
1947. Feynman$^1$ noted that the infinite integrals that were
blocking the calculations in those days were present because of the
overlapping of the delta functions in the photon or fermion
Green's functions. He and a number of others have invented forms of
renormalization to subtract the infinite terms. However, they were
unable to avoid the appearance of these infinities in the first
place. A means for avoiding the infinity in the {\it electron}
self-energy has been published$^2$. However, the renormalization
procedure for the vacuum polarization contribution to the Lamb
shift absorbs an infinite term into the electric charge. No way of
guaranteeing a {\it finite} charge from this procedure has been
suggested. A way of avoiding this problem will be shown here.

We note that Green's functions are often used to solve problems in
potential theory involving Poisson's equation. These Green's
functions often have the form, $|\hbox{\bf r}_2-\hbox{\bf r}_1
|^{-1}$. This is to be compared with Feynman's$^3$ propagator which
is,
$$\underline K_+(2,1)=i(i\rlap/\partial +m)I_+(2,1)\eqno(1)$$
where
$$I_+(2,1)=-(4\pi)^{-1}\delta (s^2) +(m/8\pi s)H_1^{(2)}(ms)
\eqno(2)$$
where $s=(t^2-{\bf x}^2)^{1/2}$ for $t^2>{\bf x}^2$ and
$s=-i({\bf x}^2-t^2)^{1/2}$ for $t^2<{\bf x}^2$. $H_1^{(2)}$ is the
Hankel function. The three dimensional Green's function used in
potential theory has no delta function while the four dimensional
Green's functions for the electron and photon do. When we consider
the vacuum polarization electron-positron loops, it will be clear
that the delta functions from the two fermion propagators will have
the same argument. This is an example of a case where an infinity
occurs.

In Section II, a new fermion Green's function will be derived in a
manner similar to the derivations often used for potential theory.
In Section III, they will be applied to obtain the formula for the
contribution of the fundamental single-loop vacuum polarization to
the Lamb shift. This formula is identical to that derived using
dimensional renormalization. Some implications of this result will
be discussed in Section IV.

II. {\bf THE FERMION GREEN'S FUNCTION}
\vskip .4cm
In order to ease the visualization of the steps required to
define a covariant Green's function, we make use of a Wick rotation
to obtain a complex time coordinate defined by $x^4=ict$ and a
$\gamma^4=i\gamma^0$ so that $\rlap/\partial=\gamma^4
\partial_4+\gamma^j \partial_j$. We define $R^2$ to equal $(x^1)^2
+(x^2)^2 +(x^3)^2 +(x^4)$ and ${\rlap{\hbox{$\sqcup$}}\sqcap}^2=
(\partial_4)^2 + \nabla^2$.

We wish to solve the equation,
$$(i\rlap/\partial_1-m)\underline\Psi(1) =i\underline S(1),
\eqno(3)$$
where the underlines indicate matrices. We require the Green's
function $\underline K(2,1)$ to be a solution of
$$\underline K(2,1)(i\gamma\cdot{\overleftarrow\partial}_1 +m)=0
\eqno(4)$$
where the arrow indicates differentiation to the left.

Our problem is to solve Eq.~(3) for $\underline \Psi$ at a point 2
given its values at points numbered 1 on a surface enclosing it and
the source $\underline S(1)$ lying inside the surface. See
Figure~1. To avoid any singularity at point 2, we introduce a small
sphere with radius $\epsilon$, centered at the point 2, and shift
point 2 to its surface. The radius, $\epsilon$, will be made
vanishingly small when all other calculations have been finished.
To obtain an integral equation, we multiply Eq.~(3) by $\underline
K(2,1)$ from the left and integrate ${\bf x}_1$ {\it over the
region exterior to the sphere containing point 2} and within the
outer surface. Then we multiply Eq.~(4) from the right by
$\underline\Psi$ and {\it again limit it to the region exterior to
the sphere and within the outer surface}. Finally we add the
altered Eq.~(4) to the altered Eq.~(3) and divide by $i$. The
result is
$$\int \underline K(2,1)\gamma\cdot({\overrightarrow\partial}_1
+{\overleftarrow\partial}_1) \underline\Psi(1)d^4x_{1} =\int
\underline K(2,1) \underline S(1) d^4x_{1}\eqno(5).$$
The left hand side of this equation is equivalent to
$$\eqalignno{\int \partial_{1\mu}\cdot
[\underline K(2,1)\gamma^\mu \underline \Psi(1)] d^4x_1
&=\int_\Sigma [\underline K(2,1) \gamma^\mu
\underline\Psi(1)] d\Sigma_{1\mu} \cr=\int_{\Sigma_0}
[\underline K(2,1) \gamma^\mu
\underline\Psi(1)]d\Sigma_{1\mu} &- \int_\sigma
[\underline K(2,1) \gamma^\mu \underline\Psi(1)]
d\sigma_{1\mu}&(6)\cr}$$
where we have made use of Gauss's theorem, and where $\Sigma$ is
the surface in four dimensions that limits the
integration. It is made up of the outer surface $\Sigma_0$ and the
infinitesimal spherical surface $\sigma$ containing the point 2.
The derivation of the surface integral from the volume integral
usually gives an {\it outward} pointing surface normal 4-vector for
$d\Sigma$ and this is the case for $\Sigma_0$, but we reverse the
{\it inward} pointing surface normal 4-vector on $\sigma$ so that
the surface normal 4-vector $d \sigma$ on this infinitesimal
spherical surface points {\it outward}. This has reversed the
sign in front of the integral over $\sigma$ in the last equation.

When the radius $\epsilon$ gets very small, the point $1$
in the integral over $\sigma$ approaches $2$ so
that $\Psi$ can be factored out of it. Replacing the left
hand side of Eq.~(5) by the right hand side of Eq.~(6) then gives
$$\int_{\Sigma_0} [\underline K(2,1) \gamma^\mu
\underline\Psi(1)]d\Sigma_{1\mu} - \Biggl[\int_\sigma
\underline K(2,1)d\sigma_{1\mu}\Biggr]
\gamma^\mu_1\underline\Psi(2) =\int \underline K(2,1)
\underline S(1) d^4x_{1}. \eqno(7)$$

In order to determine $\underline K(2,1)$, it is convenient to
write it in terms of a scalar function so that
$$\underline K(2,1)=\phi(R_{21}) (-
i\gamma\cdot{\overleftarrow\partial}_2 +m). \eqno(8)$$
Then if we multiply this equation from the right by
$(i{\gamma\cdot\overleftarrow\partial}_2+m)$, Eq.~(4) shows that
$\phi$ satisfies
$$({\rlap{\hbox{$\sqcup$}}\sqcap}^2_2-m^2)\phi(R_{21}) =0
\eqno(9)$$
This is the equation satisfied by $\phi$ in the region between the
surface $\sigma$ and the surface $\Sigma_o$. The integral over
$\sigma$ in Eq.~(7) is then, according to Eq~(8),
$$\int_\sigma \underline K(2,1) d\sigma^\mu_1=-i\gamma^\eta
\int_\sigma \partial_{1\eta}\phi(\epsilon) d\sigma^\mu_{1}
+m\phi(\epsilon)\int_\sigma d\sigma^\mu_{1} . \eqno(10)$$
The last integral vanishes because $d\sigma^\mu_{1}$ is a component
of a vector of constant length that will point in every direction
with equal weight. Since
$$\partial_{1\eta}\phi(R_{21})=x_{1\eta}\phi'(R_{21})/R_{21},
\eqno(11)$$
Eq.~(10) reduces to
$$\int_{\sigma} \underline K(2,1) d\sigma^\mu_1=-i\gamma^\eta_1
\int_{\sigma}x_{1\eta} d\sigma^\mu_{1} \phi'(\epsilon)/\epsilon .
\eqno(12)$$

Since $\sigma$ is a four-dimensional sphere, the value of the last
integral is the same for any value of $\eta$. Let us choose four
for $\eta$. The integral over the spatial dimensions will then
vanish for values of $\mu$ unequal to four. We shall use $\theta$
and $\phi$ for the
usual angles in three dimensions and add $\chi$ as the angle
between an arbitrary vector and $x^4$.
(See Ref.~4, Appendix B, for angles in four dimensions.) For a
point on $\sigma$,
$$x_4=g_{44}x^4=x^4=\epsilon\, cos\chi$$
and
$$d\sigma^4_1=\epsilon^3\, sin\theta\, sin^2\chi\, cos\chi\,
d\theta\, d\phi\, d\chi \, .$$
From these two equations, it is clear that
$$\int_\sigma x^4_1\, d\sigma^4_1=\epsilon^4\pi^2/2 \eqno(13)$$
and
$$\int_\sigma x_{1\eta}\, d\sigma^\mu_1 =g_{\eta}^\mu\,
\epsilon^4\, \pi^2/2. \eqno(14)$$
We can substitute this expression into Eq.~(12) to obtain
$$\int_\sigma {\underline K}(2,1)\, d\sigma^\mu_{1} =- i
\gamma^\mu_{1}\, \epsilon^3\, (\pi^2/2)\phi'(\epsilon). \eqno(15)$$

If we reduce Eq.~(9) to a radial equation, it will be Bessel's
equation. We need a solution $\phi$ with the property that, when
substituted into Eq.~(12) along with the value of the integral in
Eq.~(14), it will cancel $\epsilon^4$ and leave a constant. Such a
function is
$$\phi(R)=a H^{(1)}_1(imR)/R=aH^{(2)}_1(-imR)/R \eqno(16)$$
where $H^{(1)}_1$ is a Hankel function. The Bessel equation is
linear so that $a$ is an arbitrary constant. This function
$\phi(\epsilon)$ has the limiting form $-2a/(\pi m \epsilon^2)$ as
$\epsilon$ vanishes. We now substitute this into Eq.~(15) and
the result, in turn, into Eq.~(7). Then this equation can be
rearranged to give
$$\Psi(2)=-\int \underline K(2,1) S(1) d^4x_1
+\int_{\Sigma_o}\underline K(2,1) \gamma^\mu\Psi(1)
d\Sigma_{1\mu} \eqno(17)$$
where the constant $a$ has been assigned the value $im^2/8\pi$.

If we have a boundary $\Sigma_o$ in the form of an initial flat
constant time surface, a large sphere in space at intermediate
times, and a final flat constant time surface, we can undo the Wick
rotation and obtain the result,
$$\Psi(2) =\int \underline K(2,1) \underline S(1) d^4x_1 -\int_f
\underline K(2,1)\gamma^0\Psi(1)
d^3x_1+ \int_i\underline K(2,1) \gamma^o \psi(1) d^3x_1,
\eqno(18)$$
where the subscript $i$ indicates the initial and $f$ indicates the
final times. This equation agrees with Feynman's$^3$ Eq.(19) (on
his page 754)
if there is no source $\underline S$.
However, Feynman's Green's function is given in Eqs.~(1) and (2),
and ours replaces Eq.(2) by
$$I_+(2,1)=\phi(R_{21}) = (m/8\pi)H^{(2)}_1(-imR_{2,1})/R_{(2,1)}.
\eqno(19)$$
This differs from Feynman's by only a delta function. A similar
derivation can be carried out for the photon, and, again, the
Green's function has no delta function although Feynman's does.
As Feynman points out, the overlapping of these delta functions
produces many, perhaps all, of the infinities in QED. We will see
in the next section that our Green's function for fermions gives
the single loop vacuum polarization for the hydrogen lamb shift
without infinities.
\vskip .4cm
\noindent {\bf III. THE VACUUM POLARIZATION FOR THE LAMB SHIFT}
\vskip .4cm
A large number of text books present the calculation of the vacuum
polarization contribution to the Lamb shift, and we need not repeat
all of their analysis. The portion that interests us is the
single-loop contribution to $\Pi^{\mu\nu}$ that is represented by
Fig.~2. It is a loop involving an electron and a positron
propagator connecting a bare and a dressed vertex. In this case,
the excluded part of the volume is enclosed by an infinitesimal
sphere centered on one of the vertices in space-time. Its radius
$\epsilon$ is made vanishingly small at the end of the calculation.
The setting up of the expressions to be evaluated would be most
direct in configuration space-time since the inner excluded region
is a sphere in that space. However these expressions are, as
usual, most easy to evaluate in four-momentum space. For this
purpose, we can use Feynman's propagator and then subtract out the
contributions that originate from the excluded sphere. This
excludes the delta functions since they lie inside the spherical
surface.

The expression (to lowest order) to be evaluated in this manner
is$^5$
$$\Pi^{\mu\nu}(\overline q)=i4\pi \alpha\, Tr\Bigl[\int \underline
S(\overline p) \gamma^\mu\underline
S(\overline p-\overline q)\gamma^\nu d^4p/(2\pi)^4\Bigr]\eqno(20)$$
where an overline indicates a four vector, Tr indicates a trace,
and $\underline S^{-1}(\overline p)= {\rlap/p}-m$. The four-vector
$\overline q$ is the momentum of the photon connected to each of
the vertices. Our first step is to perform the indicated trace. The
resulting integral has an integrand with two nonidentical factors
in its denominators. They can be made identical with the aid of
Feynman's transforms$^4$.
The results can finally be reduced to
$$\Pi^{\mu\nu}(\overline q)={m^2\,\alpha\over 2\pi}
\int^1_{-1}\{g^{\mu\nu}[ (1 - \eta)\, M(\eta)
+{\scriptscriptstyle{1\over 2}}L(\eta)]+ (q^\mu q^\nu / q^2)\,
2\eta\, M(\eta ) \}\, d\beta. \eqno(21)$$
where
$$\eta=q^2(1-\beta^2)/4m^2, \eqno(22)$$
$$M(\eta)=M_0+M_1(\eta), \eqno(23)$$
and
$$L(\eta)=L_0+\eta L_1 + L_2(\eta). \eqno(24) $$
In addition,
$$M_0=\int^\infty_0 {u\, du \over (1+u)^2}, \eqno(25)$$
$$L_0=\int^\infty_0 {u^2\, du\over (1+u)^2}, \eqno(26)$$
$$L_1=-2\int^\infty_0 {u^2\, du\over (1+u)^3}, \eqno(27)$$
$$M_1(\eta)=\int^\infty_0\Bigl[{1\over (1+\eta +u)^2}
-{1\over (1+u)^2} \Bigr] u\, du=-ln(1+\eta) ,\eqno(28) $$
$$\eqalignno{L_2(\eta)&=\int^\infty_0 \Bigl[{1\over(1+\eta + u)^2}
-{1\over(1+u)^2}+{2\eta \over (1+u)^3}\Bigr]u^2\, du\cr
&=2\bigl[(1+\eta)\, ln(1+\eta)-\eta\bigr]. &(29)\cr}$$

The integrals $M_0$, $L_0$, and $L_1$ are independent of $\eta$
and, therefore, of $q^2$. When they are substituted into
$\Pi^{\mu\nu}$ and multiply another term that is independent of
$\eta$, the Fourier transform to space-time produces a delta
function located at the center of the spherical excluded region.
For this reason, they should not have been included in the
expression for $\Pi^{\mu\nu}$. Similarly, when $M_0$, $L_1$, or
$M_0$ multiply the $q^{2n}$ in $\eta^n$, this produces a
${\rlap{\hbox{$\sqcup$}}\sqcap}^{2n}$ operating on a delta
function. Thus when they are transformed back to space-time, they
are also confined to the center of the sphere. They, too, should
not have been included in $\Pi^{\mu\nu}$. The same is true for the
for the term $2\eta$ that appears in $L_2$. This leaves only $M_1$
in Eq(28) and the $2(1  + \eta)\, ln(1 + \eta)$ portion of $L_2$ to
be substituted into $\Pi^{\mu\nu}$. Substituting these symbols into
$\Pi^{\mu\nu}$ then gives
$$\Pi^{\mu\nu}={m^2\alpha\over 2\pi}\int_{-1}^1 \biggl\{g^{\mu\nu}
\biggl[-(1-\eta)ln(1+\eta) +[(1+\eta)ln(1+\eta) -\eta] \biggr]-
{q^{\mu}q^{\nu}\over q^2}2\eta\, ln(1+\eta)\biggr\}
d\beta.\eqno(30) $$
The $\eta$ should not be shown because it produces the derivative
of a delta function in the sphere, and the remainder of the
integrand cancels down to leave
$$\Pi^{\mu\nu}={m^2\alpha\over \pi}\int_{-1}^1\biggl(g^{\mu\nu}-
{q^{\mu}q^{\nu}\over q^2}\biggr) \eta \,\, ln(1+\eta)\, d\beta\, .
\eqno(31) $$ Since $\eta$ depends upon $\beta^2$, the integral can
be written as twice the integral from zero to one.

In order to make a comparison with the work of others, it is
convenient to define $x$ so that, from Eq.~(22),
$$\eta={q^2\over 4m^2}(1-\beta^2)= {q^2\over m^2} x(1-x).
\eqno(32)$$ from which $\beta=1-2x$. Substituting these expressions
into Eq.~(31) then gives
$$\Pi^{\mu\nu} = (q^2g^{\mu\nu}-
q^{\mu}q^{\nu})\,\pi(q^2)\eqno(33)$$
where
$$\pi(q^2)={2\alpha\over \pi}\int_0^1 x(1-x)\, ln\biggl(1+
{q^2\over m^2}x(1-x)\biggr)\, dx. \eqno(34)$$
To convert the fine structure constant $\alpha$ to $e^2$ where $e$
is the
charge associated with each vertex, we divide by $4\pi$, with the
result that our equation becomes
$$\pi(q^2) = {e^2\over 2 \pi^2}\int_0^1 x(1-x)\, ln\biggl(1+
{q^2\over m^2}x(1-x)\biggr)\, dx. \eqno(35)$$
This equation is identical to Weinberg's$^5$ Eq~(11.2.22) obtained
by using dimensional renormalization. He uses it to derive a
contribution of -27.13MHz to the Lamb shift of hydrogen. This
number is in agreement with experiment.
\vskip .4cm
\noindent{\bf IV. DISCUSSION}
\vskip .4cm
The method presented here for the evaluation of the one loop vacuum
polarization contribution to the photon self energy avoids any
infinities and, therefore, has no need for renormalization. The
same is true for the self energy diagram for the electron presented
in Reference 2. It is, therefore possible to compute the Lamb shift
for the hydrogen atom to the lowest order without renormalization.
If such finite techniques can be shown to work to all orders, it
will indicate that the equations of quantum electrodynamics are
without defects and that the presence of infinities in calculations
with them are due to faulty techniques used to solve them.
\vfill\eject
\noindent{\bf V. ACKNOWLEDGEMENTS}
\vskip .4cm
This is the third paper in a series that would not have been
possible without the mathematical techniques introduced by the late
H. S. Green.\vfill\eject

\centerline{\bf REFERENCES}
\vskip .15in
\item{1.} R. P. Feynman, Phys. Rev. {\bf 76}, 769 (1949).
\item{2.} H. S. Green, J. F. Cartier, and A. A. Broyles, Phys. Rev.
D {\bf 18}, 1102 (1978).
\item{3} R. P. Feynman, Phys. Rev. {\bf 76}, 749 (1949).
\item{4.} R. P. Feynman, {\it Quantum Electrodynamics} (W. A.
Benjamin, New York, 1962), p.139.
\item{5.} Steven Weinberg, {\it The Quantum Theory of Fields}, V.I,
(Cambridge, New York,
1995).

\vfill\eject

\psfig{file=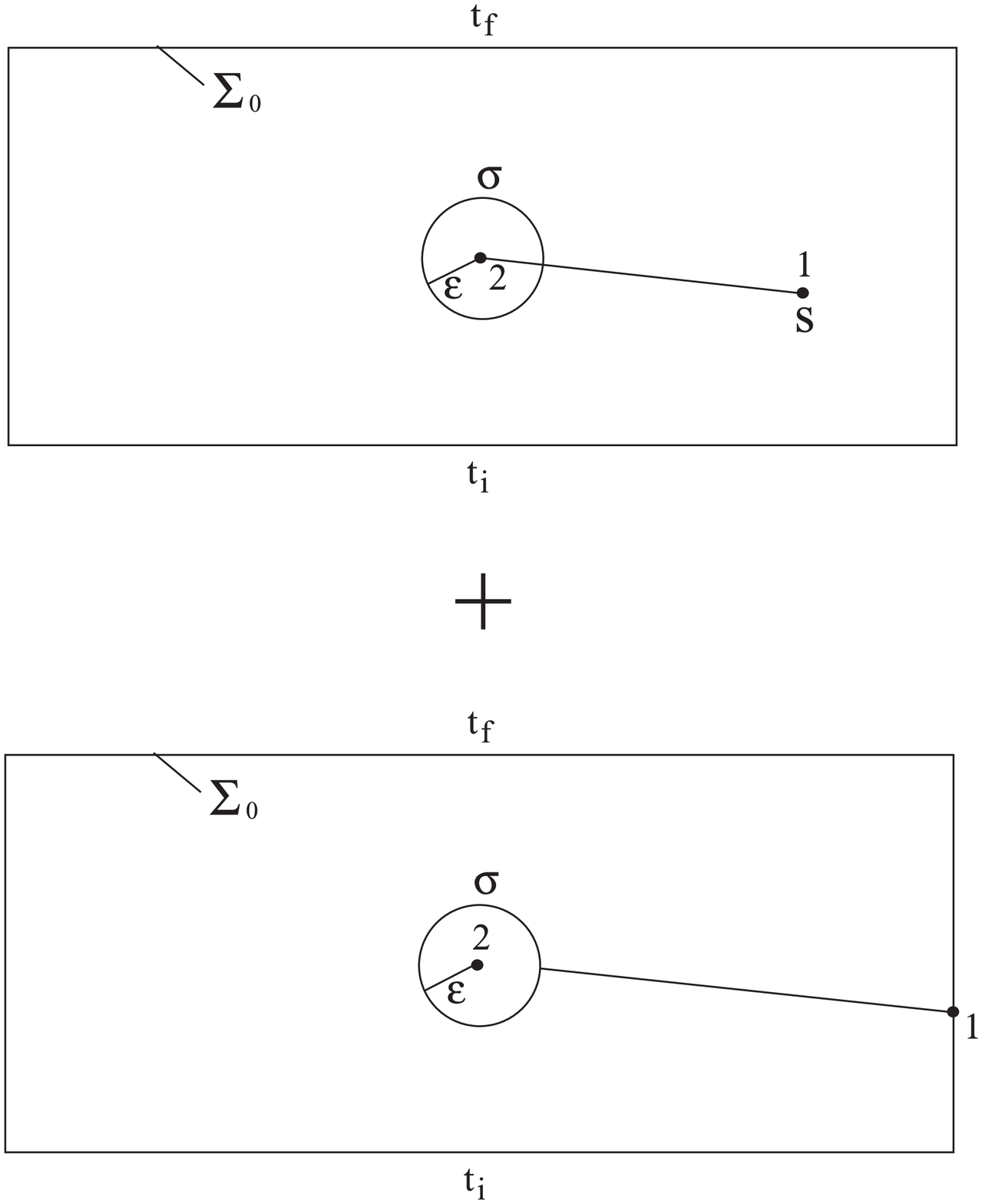,height=7.7in}
\vskip .15in
\item{\bf Figure 1} \ The outer surface $\Sigma_0$ includes $t_i$,
$t_f$, and the space cylindrical
surface.  The inner sphere $\sigma$ has a radius $\epsilon$ that
goes to zero in the end.
The point $S$ lies in the source.

\vfill\eject

\psfig{file=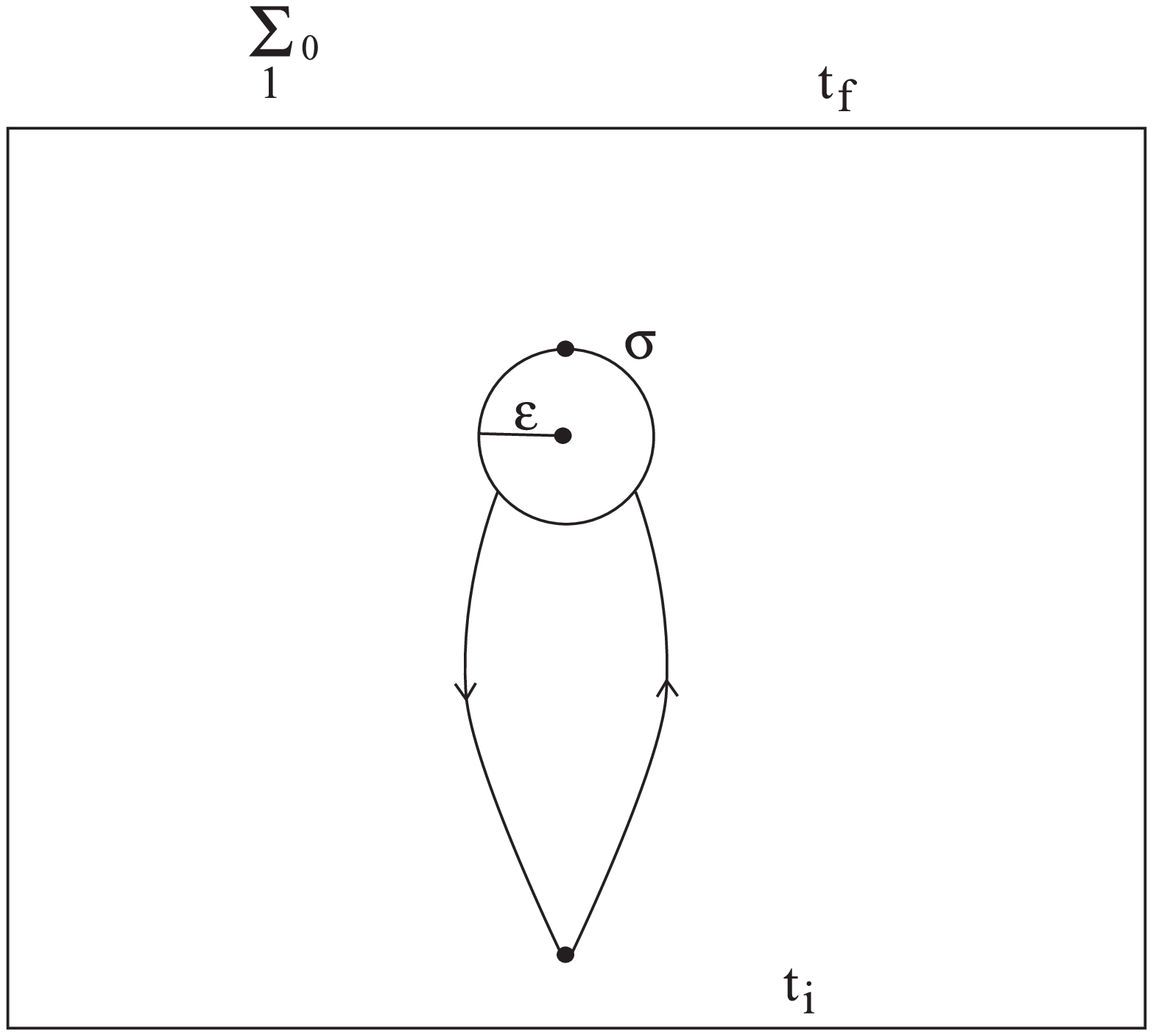,height=5.7in}
\vskip .15in
\item{\bf Figure 2} \ Vacuum Polarization Feynman diagram with
surface $\sigma$
enclosing the excluded region.

\end